# Cuddle-Fish: Exploring a Soft Floating Robot with Flapping Wings for Physical Interactions

Mingyang Xu
Keio University, Graduate School of Media Design
Yokohama, Japan
mingyang@kmd.keio.ac.jp

Jiayi Shao
Keio University, Graduate School of Media Design
Yokohama, Japan
roseshao@kmd.keio.ac.jp

Yulan Ju
Keio University, Graduate School of Media Design
Yokohama, Japan
yulan-ju@kmd.keio.ac.jp

Ximing Shen
Keio University, Graduate School of Media Design
Yokohama, Japan
ximing.shen@kmd.keio.ac.jp

Qingyuan Gao
Keio University, Graduate School of Media Design
Yokohama, Japan
girafferyeo13@gmail.com

Weijen Chen
Keio University, Graduate School of Media Design
Yokohama, Japan
weijen@kmd.keio.ac.jp

Qing Zhang
The University of Tokyo
Tokyo, Japan
qzkiyoshi@gmail.com

Yun Suen Pai
University of Auckland
Auckland, New Zealand
yun.suen.pai@auckland.ac.nz

Giulia Barbareschi
Keio University, Graduate School of Media Design
Yokohama, Japan
barbareschi@kmd.keio.ac.jp

Matthias Hoppe
Keio University, Graduate School of Media Design
Yokohama, Japan
matthias.hoppe@kmd.keio.ac.jp

Kouta Minamizawa
Keio University, Graduate School of Media Design
Yokohama, Japan
kouta@kmd.keio.ac.jp

Kai Kunze
Keio University, Graduate School of Media Design
Yokohama, Japan
kai@kmd.keio.ac.jp

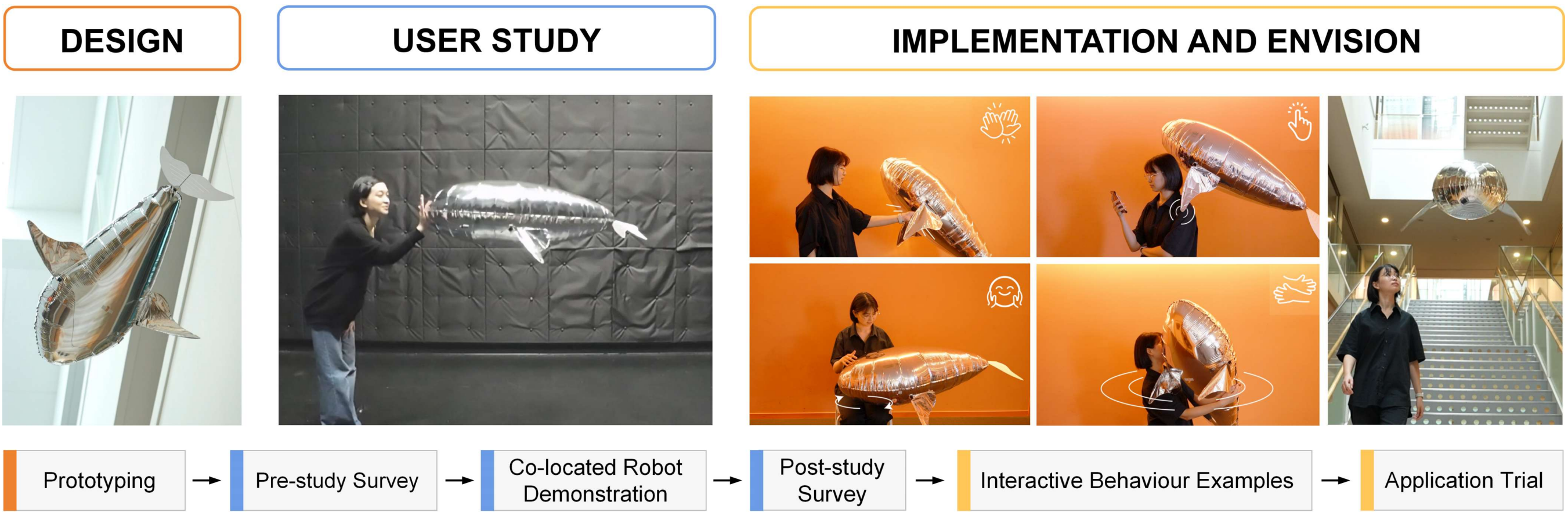

**Figure 1: This work explores how people interact with a soft floating robot in close proximity. We conducted a study that began with the robot's design and progressed to a co-located human-robot demonstration. Through participatory design sessions, participants proposed interactive behaviors, and the insights gained informed potential application scenarios.**

## Abstract

Flying robots, such as quadrotor drones, offer new possibilities for human-robot interaction but often pose safety risks due to fast-spinning propellers, rigid structures, and noise. In contrast, lighter-than-air flapping-wing robots, inspired by animal movement, offer a soft, quiet, and touch-safe alternative. Building on these advantages, we present *Cuddle-Fish*, a soft, flapping-wing floating robot designed for close-proximity interactions in indoor spaces. Through a user study with 24 participants, we explored their perceptions of the robot and experiences during a series of co-located demonstrations in which the robot moved near them. Results showed that participants felt safe, willingly engaged in touch-based interactions with the robot, and exhibited spontaneous affective behaviours, such as patting, stroking, hugging, and cheek-touching, without external prompting. They also reported positive emotional responses towards the robot. These findings suggest that the soft floating robot with flapping wings can serve as a novel and socially acceptable alternative to traditional rigid flying robots, opening new potential for applications in companionship, affective interaction, and play in everyday indoor environments.

## CCS Concepts

## Keywords

## 1 Introduction

Research on ubiquitous flying robots is gaining traction, with growing interest in applications that foster engaging human interactions [55]. Compared to ground-based alternatives, flying robots can move freely in three-dimensional (3D) space [16, 37]. This mobility opens new possibilities for human-robot interaction (HRI) that bypass spatial considerations around terrain configuration, presence of ground obstacles, and physical height differences, offering dynamic and interactive experiences [48, 59]. When designing HRI experiences where flying robots are in close proximity to people, prioritizing safety is essential. However, most existing social flying robots to date are based on quadrotor drones [5, 17, 20, 21, 24, 57], which pose safety risks due to their fast-spinning propellers [59]. Furthermore, the noise, wind, and limited flight time diminish quadrotors' suitability for sustained and close-proximity interactions with humans [27]. Thus, to leverage the advantages of flying robots for exploring novel close-contact interactions, there is a need for safer, quieter, and more approachable flying robot alternatives.

Inflatable robots offer a promising solution for safe physical interaction due to their softness and flexibility. Bioinspired flight mechanisms, such as flapping wings, present an innovative alternative to propellers. Lighter-than-air floating robots with flapping wings, such as Festo's Air_ray [19] and AirPenguin [18], mimic the movement of animals, exhibiting more organic, graceful, and potentially safer movements. The animal-like design offers opportunities to evoke social responses stemming from humans' inherent attraction to lifelike forms [56]. We argue that flapping-wing floating robots can address safety concerns while fostering meaningful and affective interactions, positioning them as ideal candidates for social robots.

While there has been limited attention on integrating such robots into human living spaces, doing so requires overcoming significant design challenges—especially in compact, lighter-than-air floating robots—since helium's lifting capacity is limited, with one litre lifting only about one gram. As a result, the floating robots are typically large in size; for example, Air_ray and AirPenguin have maximum lengths of 4.2 meters [19] and 3.7 meters [18], respectively. Their large sizes pose a significant challenge for practical use in confined indoor environments. We argue that to be suitable for indoor use in most standard houses, a floating robot should be compact enough to pass through standard door frames (typically 0.7 to 0.9 meters in width), allowing it to operate flexibly between rooms. In this paper, we present the design of a bioinspired floating robot that features simplified mechanical structures and lightweight components, and a compact wingspan of 0.78 meters. Thanks to its wing-flapping motion, the robot is capable of "swimming" in mid-air within living spaces. The robot's design aims to create a safe and inviting interaction experience for users [58].

To date, there remain gaps in understanding the user experience and design considerations for the physical interaction between humans and flapping-wing floating robots. Specifically: (1) While interaction design for rigid and ground-based robots has been extensively studied, there is a notable lack of research on physical interactions with soft and flying robots, particularly flapping-wing floating robots. The unique mobility and material properties of these robots create new interaction paradigms that remain underexplored. (2) While flapping-wing robots have been studied in aerodynamics and control systems, their role in close-proximity human interaction remains unclear. (3) There is a limited understanding of how people perceive and respond to flapping-wing floating robots. Few studies investigate users' expectations for the behaviour of these robots or explore how they can integrate into daily life in meaningful ways. To address these gaps, our study adopts an exploratory user-centred approach, focusing on human perceptions and experiences with floating robots. Specifically, we aim to answer the following research questions:

**RQ1:** What are people's perceptions of a flapping-wing floating robot operating in close proximity?
**RQ2:** Are participants willing to engage in physical interactions with a flapping-wing floating robot?
**RQ3:** What behaviours do participants envision as desirable for a flapping-wing floating robot in their daily lives?

In a user study, participants experienced six distinct co-located demonstrations and illustrated their envisioned interactions with the robot. Our results show that the presented flapping-wing floating robot is safe and attractive for close-proximity interactions. In addition to evoking positive emotions, the robot's soft and bioinspired design elicited affective behaviours like hugging and stroking, demonstrating its potential for fostering emotional connections.

Based on participants' insights, we propose two possible application scenarios for the robot: utilising the robot as a building guide and the robot acting as a playmate. We also discuss how the design of flapping-wing floating social robots, particularly those featuring soft touch and flapping-wing capabilities, has the potential to foster user engagement through interactive behaviours. Our contributions in this work can be summarised as follows:

(1) We present a soft flapping-wing floating robot, specifically designed for indoor use, emphasizing its smaller size and flexible mobility.
(2) We collected user feedback on people's perception and response (*n*=24) to a flapping-wing floating robot when it moves in close proximity, providing both quantitative and qualitative insights for researchers and designers aiming to develop socially acceptable soft flying robots.
(3) We discuss potential application scenarios and future directions for soft flapping-wing floating robots in indoor, human-inhabited spaces.

## 2 Related Works

Robots are designed to interact with humans in diverse contexts, often assuming various social roles [15, 26]. Those programmed to engage with socially meaningful cues, like body language and gestures, are known as social robots [39, 46]. Recent studies emphasize the importance of physical human-robot interaction (pHRI) for robots integrated into daily life, where safety is essential to avoid human harm [1, 44, 47]. To summarize the physical characteristics of different types of social robots, we categorize them by mobility and physical composition: **ground-based vs. flying robots** and **rigid vs. soft robots** [49], as shown in Figure 2. This taxonomy highlights each type's distinct advantages and challenges in pHRI, with our study focusing on soft flying robots.

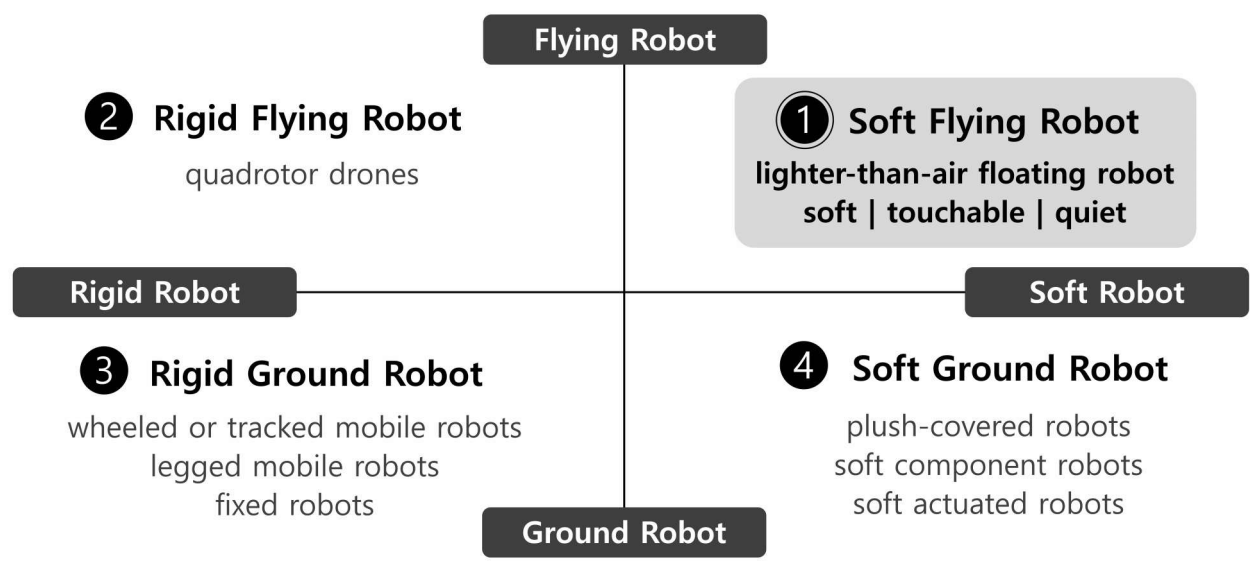

**Figure 2: Existing research reveals a gap in the study of soft flying robots within social robotics. A 2×2 matrix categorizing social robots along two key dimensions: mobility (ground-based vs. flying) and physical composition (rigid vs. soft). While conventional quadrotor drones occupy the "rigid/flying" quadrant and are often perceived as unsafe, our research explores a novel flapping-wing robot in the relatively unoccupied "soft/flying" design space as a potential alternative for close-proximity human interaction.**

The first dimension, mobility, is defined by whether the robot operates in a three-dimensional aerial environment or a terrestrial one. Ground-based robots are stable and suitable for tasks that require surface contact or manipulation. However, their movement is limited to two dimensions, restricting their operational range. Challenges such as falling, getting stuck, or failing to navigate obstacles further limit their mobility [14]. Compared to ground-based robots, flying robots are generally unaffected by terrain, like steps and stairs, and they can avoid obstacles placed on the ground [38]. Flying robots' mobility and perspective benefits enable new types of interactions through 3D movement [48, 59]. Various concepts and prototypes envision social flying robots, including jogging companions [21] and night escort drones [28]. Some are designed to dance with humans [16], provide exercise guidance [3], assist visually impaired travellers with navigation [5], and enhance users' somaesthetic experiences [29].

However, most existing social flying robots are based on quadrotor drones, which present several challenges for indoor and public use [33]. *Safety Issues*: quadrotor drones feature sharp, fast-spinning propellers that pose significant safety risks, including potential injuries from propeller strikes and accidents involving falling drones [35, 59]. Wang et al. [55] found participants in a user study expressed safety concerns, even when a small quadrotor drone was used [55]. Even with propeller guards, ensuring complete protection is challenging without compromising performance [8, 37]. *Noise and Environmental Disruption*: the high-speed rotation of propellers and motors generates significant noise and strong airflow, which can disturb users and their surroundings. Prior studies have reported that these factors can cause feelings of repugnance [55], fear [59], and increased mental stress [60]. *Psychological Safety*: even if physical safety measures such as propeller guards are implemented, users may still perceive quadrotor drones as unsafe, leading to anxiety and hesitation during interaction. Additionally, drones' tilting motions necessary for movement can also cause fear of crashes [33]. Given these limitations, multi-rotor drones are often unsuitable for enclosed public spaces or home environments [59], highlighting the need for alternative flying robot designs better suited for social interaction. Thus, further exploration is needed to investigate different form factors for user-centred designs of indoor flying robots and their pHRI capabilities in environments with close human contact.

Social robots can also be categorized as either rigid robots or soft robots. Rigid robots are generally built with hard materials, which provide greater precision and durability. Most multi-rotor drones are considered rigid robots. However, they pose a higher risk of causing harm in the event of collisions [59]. Although many soft flying robots based on soft actuators and structures have been proposed in the fields of soft robotics (e.g., [10, 11, 13, 36, 50]), in social robots, soft flying robots are primarily lighter-than-air floating robots that use helium for buoyancy. Compared to multirotor drones, floating robots hold more promise for pHRI. Inflatable envelopes serve as flexible components that act as cushions, reducing the risk of injury during contact with humans. Helium, as an inert gas, is also safe and non-flammable. Liew et al. [33] suggest that blimps or balloon drones may be a better form of companion flying robot. Additionally, floating robots often have longer flight times because they do not require continuous propeller operation to provide lift. Lighter-than-air floating robots with flapping wings are a type of soft flying robot that mimics the movements of biological organisms. Unlike traditional propeller-based systems, these robots use flapping wings for propulsion, inspired by the motion of animals. Typical examples include Festo's Air_ray [19] and AirPenguin [18]. Both designs use helium for lift and low-frequency wing flaps for thrust, allowing for smooth and flexible movement in the air. Unlike high-frequency flapping robots that mimic birds and insects [41, 61], flapping-wing floating robots do not require rapid motions to maintain lift, thanks to their helium-filled envelope. Thus, these robots only require low-frequency flapping to generate thrust. Studies have shown that people prefer lower flapping frequencies in flapping robots [43]. Additionally, with the added "wing" components, these robots may create unique interactive experiences, such as landing gently on a person's hand like a butterfly, potentially enhancing emotional connections and engagement [43].

Soft flapping-wing floating robots offer advantages in safety, mobility, and an animal-like appearance, making them promising candidates for social robots in pHRI. However, how individuals might like to interact with these social robots remains unexplored. This study investigates user perceptions and experiences during interactions to inform the design of flapping-wing floating robots, with the goal of enabling their seamless integration into everyday indoor environments.

## 3 Design of the Flapping-wing Floating Robot

Our design was driven by the goal of creating a human-friendly robot capable of safe and agile operation within typical indoor environments. To do so, we prioritised the following design considerations: (1) achieving a compact size that allows controllable flight in confined indoor spaces, (2) enabling pHRI through soft structure and gentle movement, and (3) establishing an approachable physical presentation fostering social interactions.

### 3.1 Design Form Factor

Our robot is inspired by marine creatures that move by flapping their pectoral fins. Due to the buoyancy provided by water and the lower speed requirements, marine creatures such as manta rays, sea turtles, and whales use low-frequency pectoral fin flapping to propel or assist their movement. Inspired by marine animals, we designed our flapping-wing floating robot to resemble a marine creature form, "swimming" through the air like it would in the sea. To enable pHRI, the robot is designed as an inflatable soft form with a rounded appearance [25, 52], featuring a cushioned inflatable body and soft wings. The soft envelope and wings are fabricated using aluminium-coated nylon film, making the robot's body soft and lightweight, thereby minimizing the risk of injury from collisions and making the robot suitable for physical interaction with humans.

A compact shape is necessary for floating robots to be operated effectively in indoor environments. To achieve this goal, we considered it essential for our robot to be able to pass through standard indoor doorframes, which are typically 0.7 to 0.9 meters in width. This allows the floating robot to move flexibly between different rooms, making it practical for real-world use in multi-room environments. However, reducing the width and overall size of the robot presented a unique challenge due to the limited lifting capacity of helium, which is approximately one gram per litre. Floating robots generally feature considerable helium envelopes to generate sufficient lift to match the weight, which often leads to large sizes. Air_ray and AirPenguin have wingspans of 4.20 meters [19] and 2.48 meters [18], respectively, and lengths of 2.80 meters [19] and 3.70 meters [18]. Their large dimensions pose a significant challenge for practical use in confined indoor environments. However, decreasing the size was possible only if the weight of the robot was also reduced, reducing the amount of helium available for buoyancy. To accommodate the change, we optimized the robot's propulsion mechanism, which is generally a significant contributor to the weight of floating robots. Inspired by previous research on underwater manta ray robots [12], which demonstrated that a single degree-of-freedom flapping mechanism could generate effective propulsion, we simplified our robot's flapping mechanism to a single degree of up-and-down flapping. As a result, the maximum wingspan of the robot can be reduced to 0.78 meters. Additionally, when the wings are in their lowest position, the robot's width narrows to only 0.45 meters. This means that even in tighter spaces, where doorframe widths can be less than 0.8 meters, the total width of the robot can be reduced to 0.45 meters by adjusting the wing position, allowing it to pass through the doorframe.

### 3.2 Implementation and Control

As shown in Figure 3, the main structure of the robot consists of a flapping-wing mechanism, a helium-filled envelope, a pitch-adjusting mechanism (inside the envelope), a control unit, a battery, and a tail. The robot's total weight is 70 grams. Each of the robot's two wings is powered by a lightweight 3-gram servo motor, with each motor positioned on one side of the envelope. The robot's wings are made from aluminium-coated nylon film, the same material used for the robot's envelope, and are connected to the servo motors using carbon fibre rods. The servos are controlled by an ESP32 microcontroller and powered by a 3.7V, 180mAh LiPo battery. When the wings flap at the same frequency and amplitude, the robot moves forward. To adjust the flight speed, the amplitude and frequency of the wing flaps are varied to change the thrust produced. The robot's flapping frequency can range from 0.3 to 1.5Hz, with a maximum wing amplitude of 100°. Yaw control is achieved by differential wing flapping, allowing the robot to turn by independently adjusting the speed or angle of each wing. The envelope, made from aluminium-coated nylon film, has dimensions of 1 meter in length, 0.45 meters in width, and 0.33 meters in height when fully inflated with helium. The robot's weight was carefully balanced to achieve near-neutral buoyancy, ensuring stable flight in height. The total system weight, including structural components, electronics, and power supply, is equal to or slightly greater than the buoyant force provided by the helium. To fine-tune this balance, adjustable ballast weights were incorporated near the center of mass, allowing weight modifications to compensate for component variations. Active height adjustments were achieved through body pitch control. Inside the envelope, we designed a pitch-adjusting mechanism that allows the floating robot to control its pitch by shifting its center of gravity. A servo motor inside the envelope moves an attached weight. When the centre of gravity shifts forward, the robot pitches down, and when it shifts backward, the robot pitches up. The robot is remotely controlled via a handheld transmitter operating on the 2.4GHz radio band.

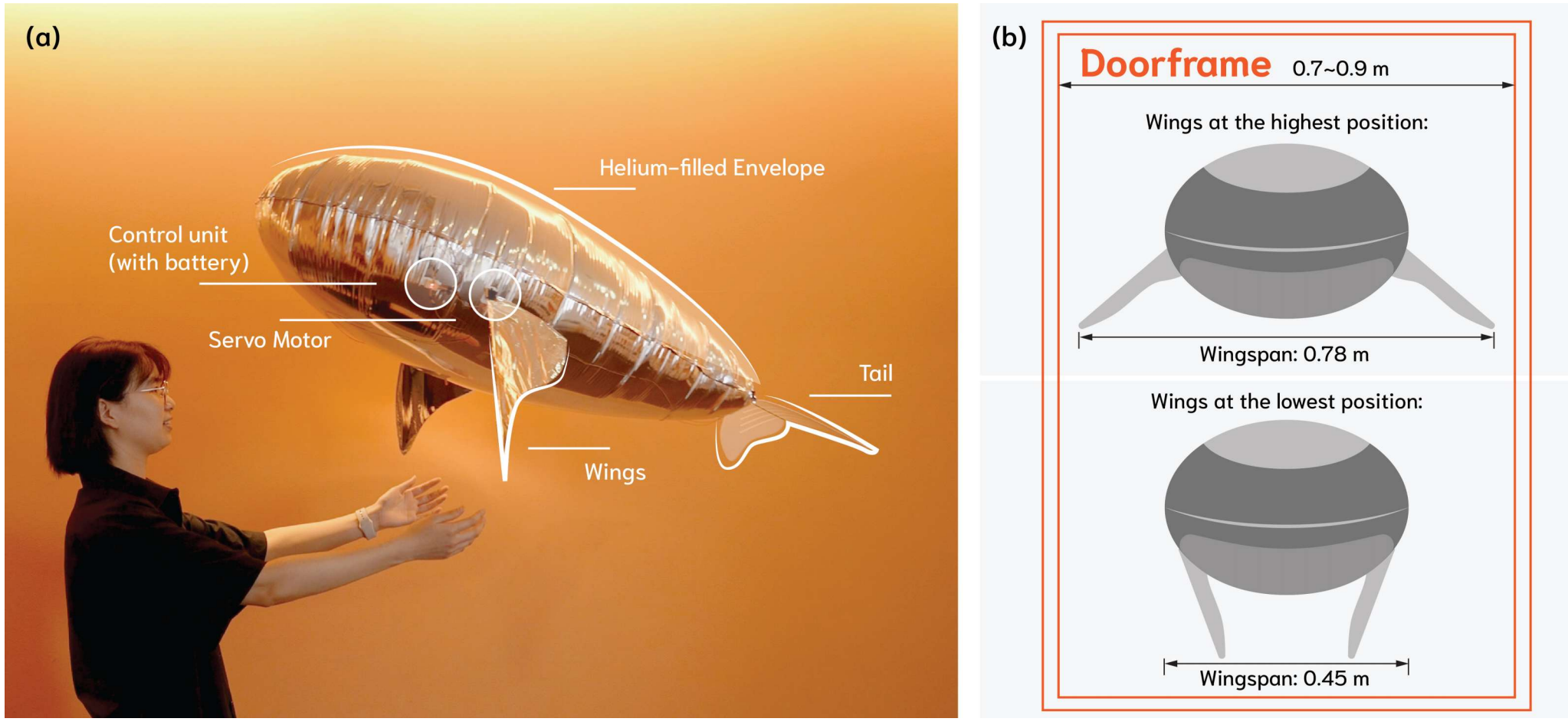

**Figure 3: (a) Key components and actuation system. The robot features a helium-filled soft body for buoyancy, two controllable wings for propulsion and maneuverability, and a tail for stability. An internal movable mass shifts the center of gravity for pitch and altitude control. The entire system is controlled by an onboard microcontroller powered by a 3.7V Li-Po battery. (b) Sized for indoor navigation. With an adjustable wingspan of 45 to 78cm, the robot can pass through standard interior doorways (typically 70 to 90cm wide), demonstrating its suitability for operation in human-centric environments.**

## 4 User Study: Evaluation of Human Perception and Envisioning of the Robot

Our user study design was inspired by previous studies in human-drone interaction [2, 8, 55] and social robot co-design [6]. We utilized a mixed-methods approach featuring both quantitative analyses of standardized questionnaires as well as qualitative analyses of behavioral observations and semi-structured interviews. To investigate people's perception and response to a flapping-wing floating robot when it moves in a co-located indoor environment (RQ1 and RQ2), we conducted a Wizard of Oz (WoZ) experiment where participants interacted with the robot remotely controlled by one of the authors through a series of pre-defined conditions. To gain deeper insights into potential applications of the flapping-wing floating robot in daily life, we invited participants to envision possible scenarios in which they would like to use and interact with the robot (RQ3).

### 4.1 Conditions

To provide participants with a basic idea of how the robot might fly and behave, we set six predefined demonstrations showcasing the robot's flight capabilities (Figure 4). In each demonstration, the robot exhibited varied movements near the participant. Each was carefully selected to elicit different forms of interaction and gauge participant reactions to various motion trajectories and degrees of proximity. **Fly Towards**: the robot flies towards the participant, approaching from the front; **Fly Overhead**: the robot flies from the front of the participant and passes overhead; **Ascend and Descend**: the robot takes off from the ground in front of the participant, ascends, and then lands back on the ground; **Circle Around**: the robot flies around the participant in a circle; **Spin Overhead**: the robot spins in circles above the participant; **Wave**: the robot pitches to achieve an upright position, then flaps one wing.

The six demonstrations were selected because they cover all of the robot prototype's basic movements, including forward motion, yaw, pitch, and wing flapping. The primary goals were twofold: first, to assess how users perceived and responded to these behaviours during interactions with the robot; and second, to inspire participants to envision more interaction scenarios with the robot. The demonstrations were implemented using Wizard-of-Oz (WoZ) methodology, with an operator piloting the robot's movements to simulate autonomy. Participants were informed that the robot was autonomous and that they could explore the robot freely in the demonstrations. No interaction prompts or instructions were provided. The robot's movements were carefully controlled to provide a consistent experience across participants. To avoid potential order effects, the sequence of demonstrations was randomized using a balanced Latin square.

### 4.2 Measurements

To capture a comprehensive spectrum of participants' emotional responses, perceptions, and experiences during the study, we utilized the following self-rating scales. **Valence and Arousal Scale:** to assess participants' emotional states before and after the co-located demonstrations (RQ1), we employed the Valence and Arousal Scale from the previous study on robotic dogs [22]. This scale is based on Russell's circumplex model of affect, emotions can be represented on two orthogonal dimensions: valence, represented as a continuum from unpleasant to pleasant, and arousal, from deactivated to activated. The dimensions are plotted on the horizontal and vertical axes, respectively. **Flying Robot Perception Scale:** we used the scale designed in [55] to measure people's perceptions of flying robots (RQ1). The scale comprises six semantic differential items: perceived safety (safe/dangerous), perceived stress (relaxed/stressful), perceived pleasantness (pleasant/annoying), perceived attractiveness (attractive/unattractive), perceived noisiness (quiet/noisy), and perceived usefulness (useful/useless). This scale employs a 7-point Likert rating, where "1" represents an extremely positive impression and "7" indicates an extremely negative impression.

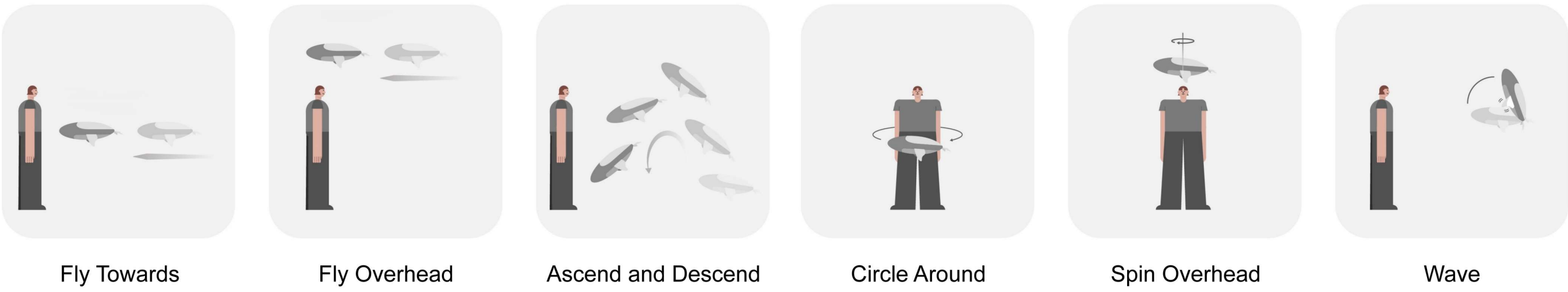

**Figure 4: The six predefined flight maneuvers used in the co-located user study were designed to investigate participants' perceptions and responses during close-proximity interactions. These maneuvers specifically included actions that, if performed by a quadrotor drone, might be perceived as threatening due to the associated risks of fast-spinning rotors, strong airflow, and loud operational noise.**

## 4.3 Participants

In total, 24 participants (15 females and 9 males) between the ages of 22 and 35 years (M = 22.67 years, SD = 3.17 years) were recruited for the laboratory-based experiment. None of the participants had experience with wing-flapping floating robots, but 6 had experience using drones. Our study was approved by the local institutional ethics review board. All participants signed the informed consent form before the experiment, no specific exclusion criteria were applied, allowing for a broad range of experiences and backgrounds. The participants were reimbursed with an equivalent of $20 after the experiment.

## 4.4 Procedure

The experiment consisted of three main phases: pre-study survey, co-located human-robot demonstrations, and post-study surveys and interviews.

*Step 1: Pre-study Survey.* Before the experiment, participants read and signed the consent form and completed a pre-study survey that collected demographic information and assessed their current emotional states using the Valence and Arousal Scale. Following this, participants were introduced to the flapping-wing floating robot and given an explanation of the study procedures.

*Step 2: Co-located Human-Robot Demonstration.* Participants were then taken to a room where they were co-located with the floating robot and experienced the six robot demonstrations. After each demonstration, participants were asked to recall and explain their responses and feelings, and then fill out the Flying Robot Perception Scale (RQ1). All studies were recorded on video and audio, with participant consent, to capture their detailed reactions and interactions (RQ2).

*Step 3: Post-study Survey.* After completing all six demonstrations, participants reported their emotional states again using the Valence and Arousal Scale. The participants' emotional changes were compared to the pre-study survey (RQ1).

*Step 4: Interactive Behaviors Design Session.* To gain deeper insights into potential applications of the flapping-wing floating robot in daily life (RQ3), we asked participants to envision specific behaviours or movements they would like the robot to perform and ways they might interact with it. To aid in ideation, we provided a design canvas inspired by [6] for social robot co-design. Following the ideation phase, participants were encouraged to handle the robot and manipulate its wings, demonstrating their envisioned behaviours to illustrate how the robot might engage with humans.

*Step 5: Semi-structured Interview.* Finally, a semi-structured one-on-one interview was conducted with each participant to gather further insights into their experience with the robot and collect comprehensive feedback and suggestions from the participants to inform the future development of the robot (RQ1). These interviews were conducted by a trained researcher using a scripted set of questions and unscripted follow-up questions based on participant responses. All interviews were audio-recorded with participant consent.

## 4.5 Data Analysis

*4.5.1 Quantitative Analysis.* In this study, quantitative analyses were conducted on the Valence and Arousal Scale to assess statistical differences between pre- and post-test measures within the same group. Prior to testing, the data were evaluated for normality using the Shapiro-Wilk test. Data that conform to normal distribution is tested by the Paired Samples t Test; otherwise, the Wilcoxon Signed-Rank Test is adopted. Additionally, quantitative analyses examined differences in participants' impressions of the floating robot across various demonstrations. Due to the violation of normality assumptions, the Friedman Test was conducted to assess group differences. For post hoc analysis, the Wilcoxon signed-rank test was applied to specific pairwise comparisons to identify significant differences.

*4.5.2 Qualitative Analysis.* The qualitative data consisted of video recordings from two camera angles (front eye-level and diagonal from behind), responses from scales, semi-structured interview transcripts, and sketches from co-design canvases. Audio recordings were transcribed using Otter AI, with the transcriptions verified by two authors against the original audio. A thematic analysis was conducted on the interview transcripts, survey responses, and video recordings. Three authors led the analysis, beginning with an initial familiarization phase followed by two rounds of coding discussions to establish a structured coding framework [9]. Two authors independently coded three video recordings, iteratively refining key themes through discussion, before one author completed coding the remaining video data. Similarly, two authors independently coded interview transcripts and survey responses, refining the codes through meetings. To ensure consistency, three authors reviewed all coded material, reaching a consensus on discrepancies. Canvas sketches were also coded separately, with recurring discussions until final themes were determined.

# 5 Results

## 5.1 Response to Floating Robot

To explore RQ2, we analyzed participants' spontaneous physical responses to the robot during the experiment based on observations from the video recordings. Among all 24 participants, 22 participants (92%) engaged in touch-based interaction with the robot, indicating that the majority were willing to engage in physical interactions with the flapping-wing floating robot. Of these, three participants (13%) touched the robot's flapping wings, whereas others made contact with the body. Additionally, 4 participants (17%) engaged in gesture-based interactions, while another 4 participants (17%) exhibited avoidance behaviours. We further summarized participant behaviours in both touch-based and gesture-based interactions in Table 1. Examples of different spontaneous interactions observed during the user study are presented in Figure 5.

## 5.2 Quantitative Results

*5.2.1 Valence and Arousal Scale.* To explore RQ1, we investigated participants' perceptions of the robot from an emotional perspective. A paired samples t-test was conducted to compare Arousal scores before and after the experiment. There was no statistically significant difference in Arousal scores before (M = -0.142, SD = 0.977) and after (M = 0.158, SD = 0.862) the experiment, t(23)=-1.331, $p = 0.196$. The mean difference was -0.300 (SD = 1.104), indicating a small, non-significant decrease in Arousal scores following the experiment. For Valence scores, a Wilcoxon signed-rank test was conducted to evaluate the differences before and after participation in the experiment. The analysis revealed a statistically significant increase post-participation, Z=-3.76, $p < .001$. Specifically, out of 24 participants, 3 showed higher Valence scores before participation, while 21 participants exhibited higher scores after participation. The median Valence score increased from 0.75 to 1.55, indicating that participation in the experiment was associated with a significant improvement.

*5.2.2 Flying Robot Perception Scale.* We investigated participants' perceptions of the robot by using a flying robot perception scale [55] to answer RQ1. As shown in Figure 6, we visualized the original ratings data on a 7-point scale using stacked bar charts to illustrate the distribution of participants' responses. Prior to testing, the data were evaluated for normality using the Shapiro-Wilk test. As the assumption of normality was violated, we performed Friedman Test to test for differences between groups. For the post hoc test, Wilcoxon signed-rank test was selected on the different combinations of related groups to examine where the differences occurred. Given the multiple comparisons, Bonferroni adjustments were applied to the significance level to control for Type I error. The original significance level of 0.05 was divided by the number of tests (15), resulting in an adjusted significance level of 0.0033. Therefore, only p-values less than 0.0033 indicate statistically significant results.

The Friedman test indicated a statistically significant difference in perceived attractiveness depending on movement type, $\chi^2(2) = 12.992$, $p = 0.023$. Post hoc Wilcoxon signed-rank tests, with the adjusted significance threshold ($p < 0.003$), revealed no significant differences among the movement types, despite observed reductions in perceived effort across trials. Median (IQR) perceived effort levels were as follows: Fly Towards = 1.5 (1 to 2), Fly Overhead = 2 (1 to 2.75), Wave = 2 (1 to 3), Ascend and Descend = 1 (1 to 2), Spin Overhead = 1.5 (1 to 2), and Circle Around = 1 (1 to 2). A second Friedman test also confirmed a statistically significant difference in perceived attractiveness depending on movement type, $\chi^2(2) = 23.415$, $p < .001$. Post hoc Wilcoxon signed-rank tests revealed significant pairwise differences: Fly Towards vs. Spin Overhead (Z = -3.593, $p < .001$), Fly Overhead vs. Spin Overhead (Z = -3.135, $p = 0.002$), and Ascend Descend vs. Spin Overhead (Z = -3.361, $p < .001$). Median (IQR) perceived effort levels for each movement type were as follows: Fly Towards = 2.0 (1.0 to 5.0), Fly Overhead = 2.5 (1.0 to 6.0), Wave = 3.5 (1.0 to 6.0), Ascend Descend = 3.0 (1.0 to 6.0), Spin Overhead = 5.0 (2.0 to 6.0), and Circle Around = 4.0 (1.0 to 6.0).

## 5.3 Qualitative Results

*5.3.1 Movement Features Shaping Interpretations.* Participants were captivated by the robot's aquatic, animal-like appearance, often associating it with creatures like dolphins, birds, or even pets. Many participants ascribed personality traits to the robot, interpreting it as a creature with emotions or intentions. For instance, P8 envisioned a "cute dolphin smiling," while P7 likened its movement to "flying like a bird" and experiencing freedom. These interpretations suggest that the robot's smooth, organic motion evokes positive personality associations, deepening the participants' emotional connection with it. Although all participants superimposed some degree of characterization on the robot, specific details varied depending on whether participants perceived the robot as a pet-like companion or a fantastical creature. P10, for example, felt a sense of warmth and nostalgia, likening the robot's behaviour to that of a beloved pet. Others saw it as a fantastical underwater creature, experiencing a sense of wonder as the robot's movement evoked memories of "whales swimming in the galaxy" (P5) or flying fish, transporting them into a dreamlike experience.

Specific robot movements also shaped participants' interpretations. Movements such as "Fly Towards" or "Circle Around" were typically interpreted as playful, similar to a pet's behaviour (P5, P7, P8, P10, P14, P15, P16, P17, P18, P23); 67% of participants responded by reaching out to the robot and making contact. In contrast, movements like "Fly Overhead" or "Ascend and Descend" evoked aquatic imagery, which participants described as "swimming" or "surfacing." These interactions mirror natural human responses to animal behaviours, with participants often describing their experience as calming and enchanting. As P15 summarised, "I felt charmed by it, like an animal can charm you—a cute animal." The robot's design and movements effectively fostered a soothing and engaging environment, prompting positive responses. In post-experiment interviews, participants shared their emotional experiences with the robot. Out of 24 participants, 20 (83.3%) said they felt calm or relaxed after experiencing it. Many explained that the robot's smooth movements reminded them of peaceful places like aquariums or the deep sea. For example, P15 said, "It felt like being in an aquarium, watching something calm." Similarly, P9 described, "The gentle flapping of its wings was like a fish swimming gracefully underwater, made me feel relaxed." These responses highlight how the robot's design and movements can create a calming and peaceful atmosphere, bringing to mind images of tranquil, natural settings. However, likening the robot to an aquatic animal, especially a large one, could at times also evoke feelings of weariness, as seen in participants who engaged in avoidance behaviours in response to the movement of the robot. For example, when witnessing the "Fly Towards", P18 reported "I felt like it was coming towards me a little bit aggressively, so I moved out of the way. I felt like I was in the ocean, and if an actual whale was coming at me, I would really bend back a little and pass."

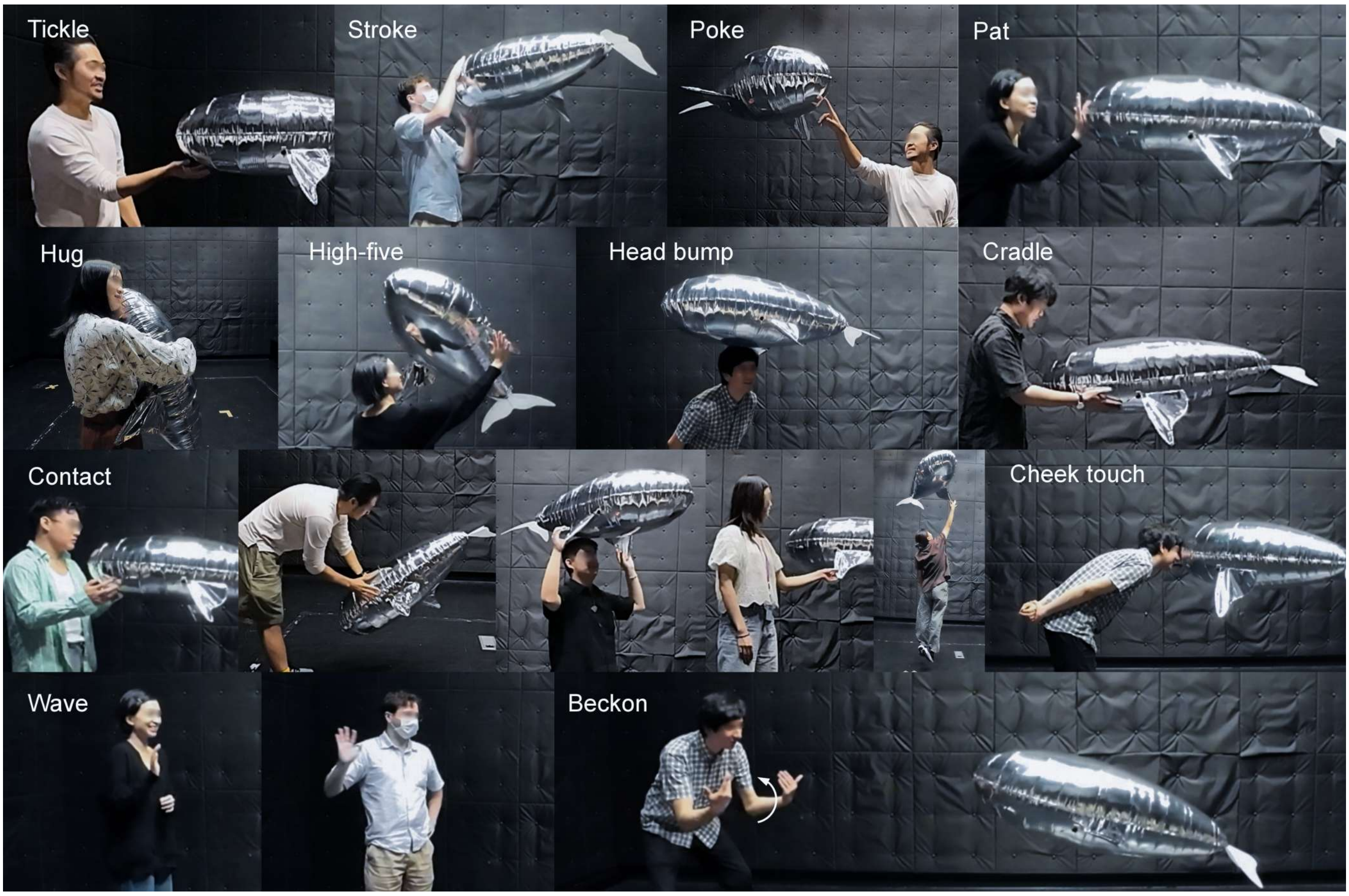

**Figure 5: Participants were free to respond to and interact with the robot in any way they felt comfortable. Examples of spontaneous responses and interactions were observed throughout the study, highlighting participants' natural engagement.**

*5.3.2 Affective Connections With The Robot.* Participants commonly attributed intentionality and personality to the robot, interpreting its movements as greetings, playfulness, or curiosity. For instance, P1 felt the robot "seemed to want some physical contact with me" as it flew close, while P18 saw it as "going on a stroll without a destination," interpreting overhead movements as calm and animal-like. The interactions where the robot circled or approached participants were often seen as gestures of companionship. Through these behaviours, participants assigned personalities to the robot, describing it as "friendly," "playful," "curious," "cute," and sometimes "clumsy" or "independent." We also capture several instances of participants' affective connections with the robot. P11, for instance, saw the robot as "complaining" when it bumped into them, leading to a feeling of wanting to engage. P21 viewed the robot's actions as efforts to "show off its abilities, like a little baby," which fostered a sense of pride and encouragement. Physical gestures like patting the robot's "head" were common as participants sought to comfort it. P8 likened it to a "sidekick," while P4 contrasted it with drones, saying, "Drones with blades feel rigid and cold, clearly just a machine or tool. But this floating robot has a rounder appearance and moves more gently. Compared to a drone, if mine broke, I probably wouldn't care as much; I would just get a new one. But if this floating robot broke, I would feel heartbroken." Such responses highlight the potential of how design elements, including the friendly appearance and relatable behaviours, can foster emotional responses and transform the robot from a technological object into a playful companion.

**Table 1: The interactions initiated by participants during the study were categorised into 12 types and classified into two interaction forms: touch-based and gesture-based interactions.**

| Interaction | | Description | Participant |
|---|---|---|---|
| **Touch-based interaction** | | | |
| Contact | 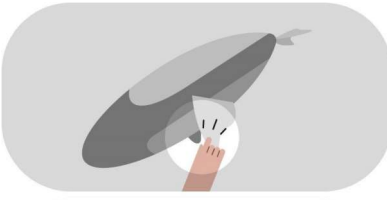 | Any touch contact with the robot that has no movement | P1, P2, P3, P4, P5, P7, P8, P10, P11, P12, P13, P14, P15, P16, P17, P19, P20, P21, P22, P23, P24 |
| Cradle | 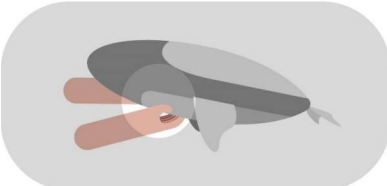 | Hold the robot gently and protectively | P4, P8, P10, P14, P21, P22, P23, P24 |
| Stroke | 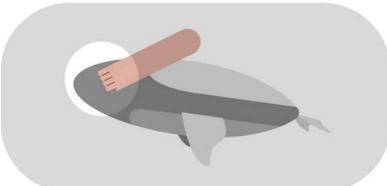 | Move hand gently over the top of the robot in a certain direction, often repeatedly | P2, P3, P7, P11, P12, P14, P16, P19, P21 |
| Poke | 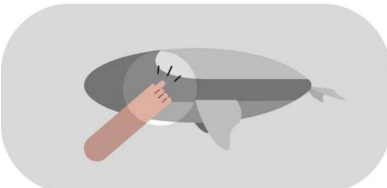 | Lightly touch the robot through one fingertip | P3, P10, P13, P15 |
| Pat | 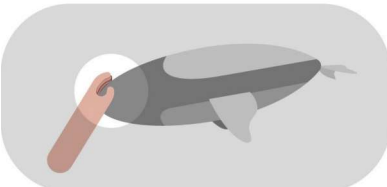 | Touch the robot using the flat part of the hand in a soft, quick motion | P2, P3, P4, P20 |
| Hug | 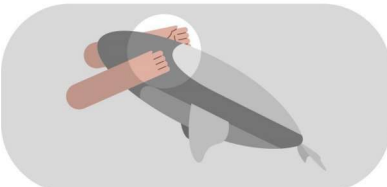 | Hold the robot closely or tightly | P7, P10, P21 |
| Cheek touch | 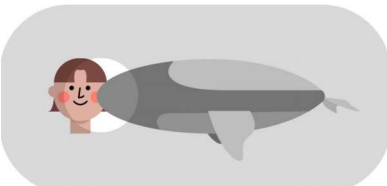 | Touch the robot lightly with the cheek | P20, P24 |
| Tickle | 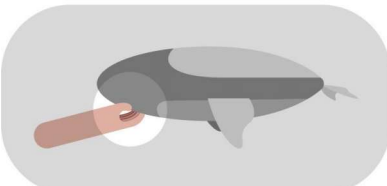 | Rub the robot using gentle movement of fingertips | P3, P8 |
| High-five | 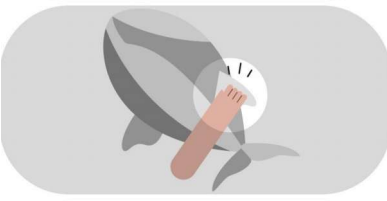 | Tap the robot's wing with hand | P20 |
| Head bump | 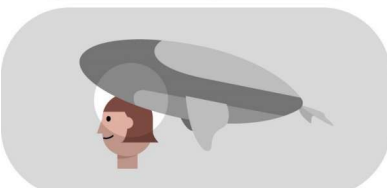 | Use the head to gently push the robot upward | P20, P24 |
| **Gesture-based interaction** | | | |
| Wave | 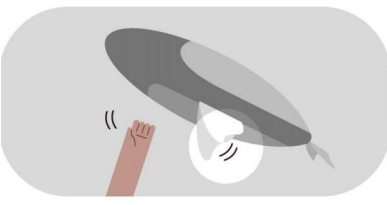 | Greet the robot with a wave of the hand | P12, P20, P23 |
| Beckon | 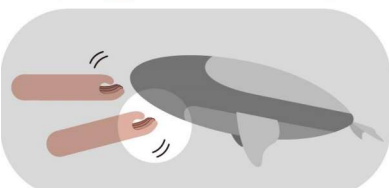 | Make a quick and repetitive motion with the hand to signal the robot to come closer | P24 |

*5.3.3 Envisioning Everyday Life with the Floating Robot.* Participants expressed a desire for the robot to fulfil various companionship roles, emphasizing its potential for emotional support. They envisioned future robots of similar appearance that can provide gentle hugs and a comforting presence. For example, P8 described a robot that reacts to its owner's presence by flying around or making sounds just for them, akin to a dog greeting their owner. P20 and P13 highlighted the robot's potential to alleviate loneliness in public spaces, suggesting roles like cheering users on during chores or floating around workplaces (see (a) in Figure 7). These envisioned interactions reflect a cultural need for accessible social engagement amid rising single-person households and social isolation.

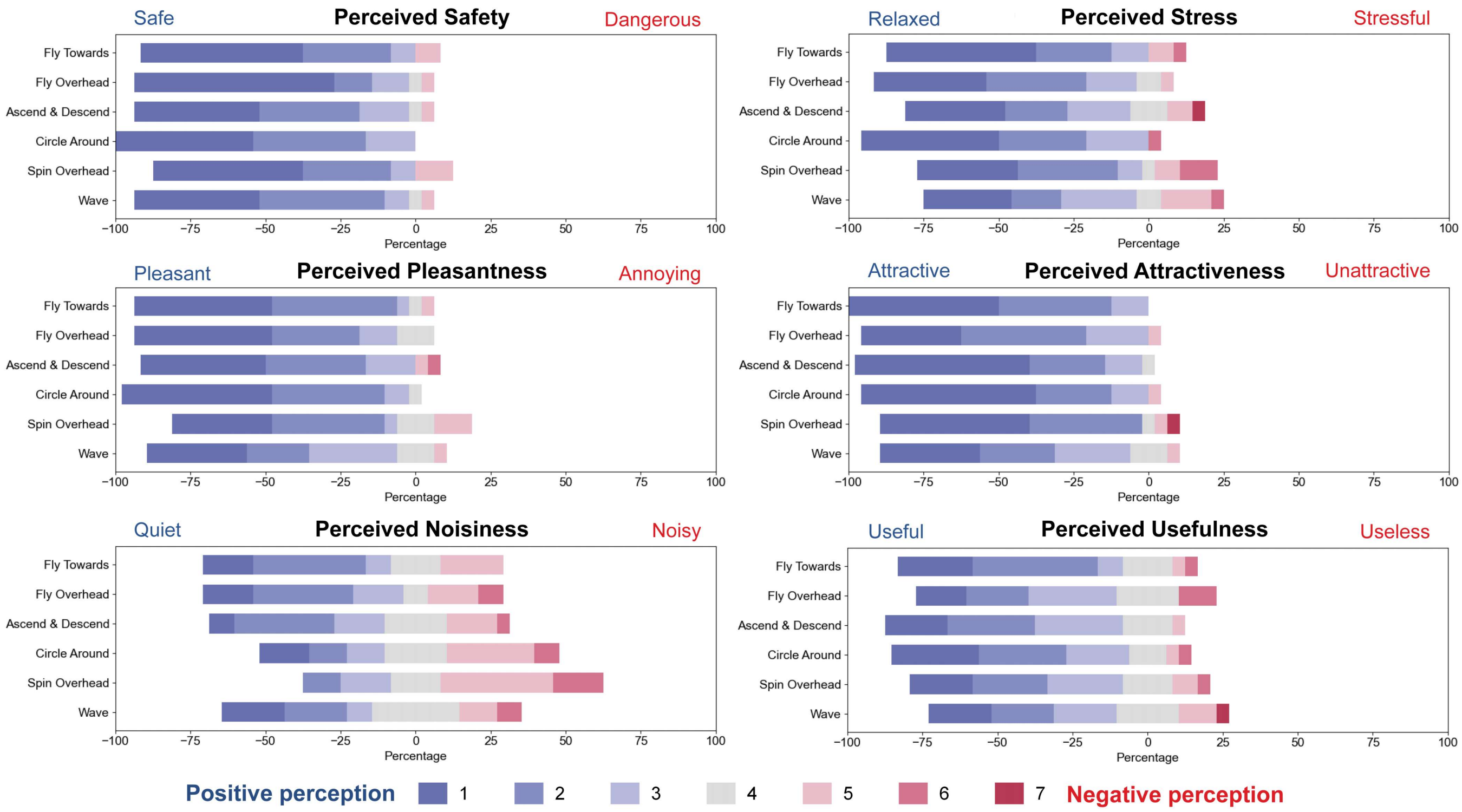

**Figure 6: Stacked bar plots illustrating participants' perceptions of the robot following six co-located demonstrations (Fly Towards, Fly Overhead, Ascend and Descend, Circle Around, Spin Overhead, and Wave). Perceptions were measured through a questionnaire with six semantic differential items on a 7-point scale: Quiet/Noisy, Pleasant/Annoying, Useful/Useless, Relaxed/Stressful, Attractive/Unattractive, and Safe/Dangerous. A score of "1" indicates an extremely positive perception (e.g., extremely quiet for Quiet/Noisy), while a score of "7" indicates an extremely negative perception (e.g., extremely noisy for Quiet/Noisy).**

Beyond companionship, participants creatively conceptualized playful roles for the robot. P5 imagined a "dance party" scenario where the robot would fly around a disco ball, encouraging fun and exercise. Suggestions included playful interactions reminiscent of pet play, such as a "poke" response or a "fetch" function, which enhances engagement through tactile feedback. Participants also valued the robot's customizability, envisioning it as a canvas for personal expression and creativity (P23) (see (b) in Figure 7). Lastly, participants sketched household-assistive interactions, emphasizing the robot's potential to alleviate daily task stress and enhance well-being. Ideas included the robot remaindering cooking ingredients or providing gentle cues to wake users calmly. Many viewed the support of the robot during chores as a way to make repetitive tasks more enjoyable (P9) while saving energy and time (see (c) in Figure 7).

Based on participants' insights, following the experiment, we trialled two possible application scenarios with an operator controlling the robot: utilising the robot as a building guide and the robot acting as a playmate. Although these application trials were conducted via remote piloting, they underscored the robot's potential to perform autonomously or semi-autonomously in future implementations.

**Guide:** The robot was deployed in a complex building where it guided a user from the entrance to a designated room, as shown (f) in Figure 1, the robot was positioned to the left and ahead of the user. Notably, when passing through stairs, the floating robot showcased the ability to fly over staircases, providing a fluid guide for the user. This advantage highlights the potential of the robot in spaces with multi-level layouts where mobility challenges restrict ground robots.

**Playmate:** We encouraged a user to engage freely with the robot. The user proposed several interaction modes: (1) following the user, (2) mirroring the user's arm movements with its wing to perform exercises together, (3) spinning together in a circular motion. This trial received positive feedback, as the user found the robot engaging and capable of creating a sense of companionship. The flapping-wing design, combined with its floating capability, added a unique and charming element to these interactions, fostering an experience that was both familiar and novel.

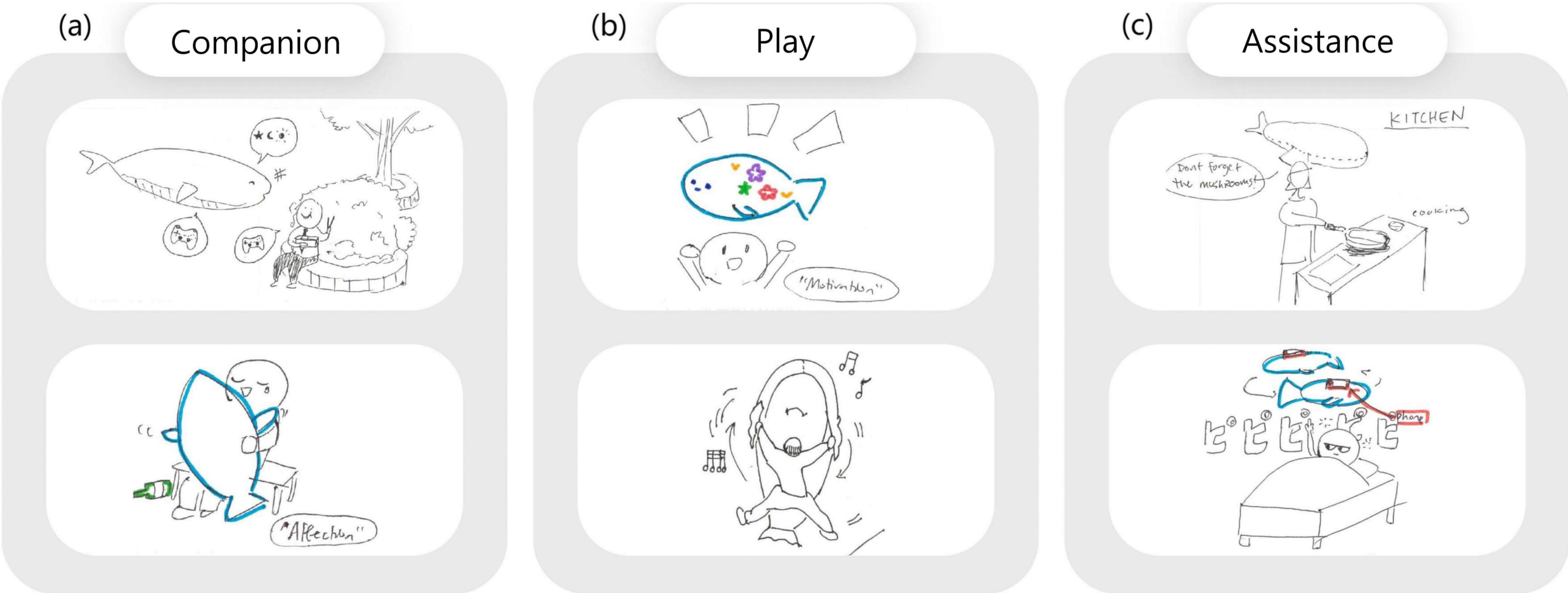

**Figure 7: Examples of participants' sketches illustrating their envisioned roles and interactions for the robot. After experiencing the six co-located demonstrations (see Figure 4), participants were invited to envision and sketch potential scenarios and interactions. Our analysis grouped these into three themes: (a) companion, (b) play, and (c) assistance.**

## 6 Discussion

To our knowledge, our study is the first to propose the design of a compact floating robot featuring wing-flapping mechanisms for propulsion, which is suitable for use in home-like environments and specifically investigates participants' experiences during the interactions and their expectations for real-world uses. Addressing **RQ1**, our findings highlight how the behavioural responses and subjective perceptions of participants are shaped by the movement of the robots, including the influence of associations made with aquatic animals and fantastical creatures. Quantitative results, such as the significant increase in valence scores and favourable perceptions captured by the Flying Robot Perception Scale, demonstrate that participants viewed the robot positively, associating it with safety, pleasantness, and attractiveness. For **RQ2**, the high percentage of participants (92%) engaging in touch-based interaction indicates the willingness to physically interact with the robot, highlighting its potential to foster tactile engagement. For **RQ3**, participants envisioned the robot in diverse roles, including emotional companionship, playful interactions, and assistive tasks, underscoring its potential to enhance everyday life by addressing both emotional and practical needs. These insights form the foundation for designing future iterations of floating robots that are expressive, interactive, and user-centred.

### 6.1 Wings as "Hands": Enabling Affectionate Gestures in Close Interaction

Developing intuitive gesture sets tailored to user preferences enhances interaction effectiveness. A systematic literature review [53] explored how gesture recognition systems can be optimized for better user engagement. Key findings emphasized the importance of involving users in the design process to ensure gestures are natural and intuitive. Our experimental design, which included co-creation sessions, allowed participants to express their thoughts on companionship with the robot, its appearance, sounds, and helpful functionalities in their everyday lives. Participants revealed individual preferences for gestures that felt organic and aligned with their preferred gestures, echoing our findings on familiar robot movements. They showed positive reactions to movements like "Fly Towards" and "Circle Around" while expressing confusion about less familiar actions like "Spin Overhead" and "Wave." This highlights the need for users to have a clear understanding of robot behaviours to foster meaningful connections.

Our study also proposed a novel communicative gesture system using the robot's fins. Previous research has extensively investigated the use of wings or fins for robotic mobility [4, 7, 34, 54]. In our study, participants expressed a desire for the robot to use its flapping wings for affectionate gestures, such as hugging them. This inspires us to liken the robot's wings to "hands" which can be used for a variety of soft social touch interactions. Integrating social touch into robots' arms and hands has been explored in several studies [23, 30, 31], illustrating the importance of tactile and social touch capabilities in enhancing human interaction. Our research stands apart by using wings for social interaction and affective communication, aligning with the emerging interest in bio-inspired designs that mimic natural behaviours.

### 6.2 Soft Touch, Positive Engagement

Designing socially comfortable robots is a key concern for making robots pervasive. A key aspect of promoting positive emotional connections is to carefully design tactile sensations since touch influences perceived safety, comfort, and emotional engagement. Research indicates that robot-initiated touch can reduce physiological stress and increase perceived intimacy, fostering stronger human-robot bonds [42]. Our experiments revealed that participants naturally engaged in gentle physical interactions, such as hugging and tapping the robot, highlighting the social and emotional connections formed with it. On the other hand, the design of flying robots poses unique challenges due to their lightweight materials, limiting the texture options. This limitation suggests a need for encouraging light, casual touch interactions rather than deep, close-contact gestures [45]. Furthermore, insights from recent studies emphasize the importance of maintaining socially comfortable distances in interactions. Cautiously approaching sensitive areas can enhance user comfort and deepen the emotional resonance for human beings, particularly in social and therapeutic contexts. These findings pave the way for future research into nonverbal and touch-based interactions [40].

## 6.3 Potential Applications

We envision integrating the ubiquitous flapping-wing floating robot into people's daily indoor environments. As an autonomous agent, the robot could function as a companion pet within the indoor space. Leveraging the mid-air space typically underutilized in home indoor environments, the floating robot can navigate stairs to move between floors, travel between rooms, and float by ceilings when inactive, thus preserving floor space and reducing clutter. This mid-air space utility offers practical advantages over ground-based devices, expanding the home environment's interaction zones without competing for space. In our study, the flapping-wing floating robot was controlled by an operator, which led us to the speculation towards a potential use as a telepresence avatar [32, 51], offering an innovative means for remote social connection and emotional engagement. As a soft telepresence avatar, the robot can be remotely operated by a person, allowing individuals in distant locations to interact with others, mitigating feelings of loneliness and fostering emotional bonds. For example, a parent working overseas could remotely control the floating robot to play with their child at home, where the child might chase or jump to touch the floating avatar, mimicking a lively game of chase as though the parent were physically present. This interaction transcends traditional screen-based communication, bringing a new sense of closeness and physicality to remote relationships. While the robot currently faces payload limitations, there are several options for mitigation. Advances in micro-drone technologies are providing increasingly lightweight and compact electronic components, broadening the range of devices suitable for integration. Additionally, necessary systems do not need to be onboard the robot. Large processing units can be stationed on the ground as a base station, transmitting data to the robot for execution, rather than relying on fully onboard processing. These strategies could enable the development of lightweight floating robots capable of supporting broader applications.

## 6.4 Limitations and Future Work

This study presents a flapping-wing floating robot that offers advantages in terms of safety and soft construction; however, it also introduces certain limitations. Due to its inflatable nature, the robot is sensitive to external environmental factors, particularly airflow, making outdoor applications challenging. Therefore, we explored its intended use in indoor environments, where its "softness" attribute is also well-suited. Even in indoor settings, however, airflow from air conditioning and fans may interfere with the robot's operation. Currently, the robot requires remote control by an operator, which limits its autonomy and adaptability. Future work will focus on developing autonomous control mechanisms that enable the robot to respond in real time to the dynamic environment. The robot's envelope is made of aluminium-coated nylon film to reduce helium leakage, but over extended periods, a helium supplement is still necessary. While the envelope remained undamaged in our user study, real-world applications will require increased durability.

Additionally, this study did not systematically investigate the influence of specific design elements—such as the form factor, flapping wing motion, and bioinspired aesthetics—on user experience. Future research explores how these elements influence user perception and response, providing insights for optimizing the robot's design. In this study, the robot's demonstrations were controlled by a pilot using the WoZ method. This resulted in slight variations in the demonstrations experienced by each participant. These variations primarily resulted from subtle differences in manual control execution. Although these variations did not significantly alter the core interaction experience, they may have contributed to individual differences in user perception. Future implementations will benefit from standardized autonomous control to ensure uniform interaction experiences across all participants. Moreover, while the participants in this study were aged between 22 and 35, the robot shows promise for applications across a broader range of age groups, including older adults and children. Future research will aim to encompass a wider demographic spectrum.

# 7 Conclusion

In this paper, we explore a soft flapping-wing floating robot capable of physical human-robot interaction. The robot features a soft envelope and achieves propulsion through flapping flexible wings. This study examines how people perceive and respond to a flapping-wing floating robot in close proximity. Through a user study, we explored their experiences with six distinct flight demonstrations indoors. Participants engaged in 10 types of touch-based interactions and 2 types of gesture-based interactions with the robot. Quantitative and qualitative results indicated positive perceptions of the robot's perceived safety. Participants also reported positive attitudes and creative visions for future interactions with floating robots in daily life. We then discuss design considerations and future application scenarios for floating robots. Our work highlights the promising opportunities for future flapping-wing floating robots to engage with humans in indoor environments.

# Acknowledgments

We sincerely thank Yunkai Qi, Xiaoyue Zhang, and Wanhe An for their work on the early prototypes. Our appreciation also goes to Yanheng Li for her help with image anonymization and to Christopher Changmok Kim for assistance with photography. Furthermore, we are thankful for the discussions with Shiqi Sun, Kanyu Chen, Sophia Sakel, and Dr. Takatoshi Yoshida. This work was supported by the JST Moonshot R&D Program "Cybernetic being" Project (Grant number: JPMJMS2013), JST Presto (JPMJPR2132), and the Keio University Doctorate Student Grant-in-Aid Program from Ushioda Memorial Fund. Additionally, this research is partially funded by the University of Auckland Faculty of Science Research Development (Grant number: 3731533) and Japan Society for the Promotion of Science (JSPS).